\documentclass[textwidth=19cm,paperheight=24cm,fleqn,10pt]{article}
\usepackage[utf8]{inputenc}
\usepackage[T1]{fontenc}
\usepackage{authblk}
\usepackage{amsmath}
\usepackage{amstext}
\usepackage{amssymb}
\usepackage{color}
\usepackage{graphicx}
\usepackage{algpseudocode}
\title{The self-organised, non-equilibrium dynamics of spontaneous cancerous buds}
\author[1]{Abramo Agosti}
\author[1,2]{Stefano Marchesi}
\author[2,3]{Giorgio Scita}
\author[1,*]{Pasquale Ciarletta}
\affil[1]{MOX Laboratory, Dipartimento di Matematica, Politecnico di Milano, piazza Leonardo da Vinci 32, 20133 Milano, Italy}
\affil[2]{IFOM, the FIRC Institute of Molecular Oncology, Via Adamello 16, 20139 Milan, Italy}
\affil[3]{Department of Oncology and Hemato-Oncology, University of Milan, Via Festa del Perdono 7, 20122 Milan, Italy}

\affil[*]{Corresponding author: pasquale.ciarletta@polimi.it}
\date{} 

\begin{document}
\maketitle
\begin{abstract}
Tissue self-organization into defined and well-controlled three-dimensional structures is essential during development for the generation of organs. A similar, but highly deranged process might also occur during the aberrant growth of cancers, which frequently display a loss of the orderly structures of the tissue of origin, but retain a multicellular organization in the form of spheroids, strands, and buds. The latter structures are often seen when tumors masses switch to an invasive behavior into surrounding tissues. However, the general physical principles governing the self-organized  architectures of tumor cell populations remain by and large unclear.
In this work, we perform  in-vitro experiments { to characterize the growth properties of glioblastoma budding emerging from monolayers}. { Using a theoretical model and numerical tools here we find that such a topological transition} is a self-organised, non-equilibrium phenomenon driven by  the trade--off of mechanical forces and physical interactions exerted at cell-cell and cell-{{ substrate}} adhesions. Notably, the unstable disorder states of uncontrolled cellular proliferation macroscopically emerge  as complex spatio--temporal patterns that evolve statistically correlated by a universal law.
\end{abstract}

%
%
\thispagestyle{empty}



Cancer cells display abnormal interactions with their surrounding micro-environment, that protect them from the attacks of the immune systems, sustain uncontrolled proliferation and promote tumour progression and even dissemination to distal site \cite{kai2016force}. Thus, understanding how mechanical and physical cues influence the invasive strategies of a malignant tissue is crucial for designing unconventional anti-tumorigenic treatments and to improve therapeutic outcomes \cite{bissell2001putting,jain2010delivering,jain2014role}.
This unmet need has triggered numerous attempts to develop suitable system models capable of reproducing not only the complex collective behavior of cancer cells, such as migration, proliferation, and resistance to therapies, but also the dynamic feedback between cancer tissues and their surroundings, recently termed mechano-reciprocity \cite{van2018mechanoreciprocity}.
Two-dimensional (2D) cultures have traditionally represented a simple and low-cost platform. Using 2D monolayers, for example, has proved to be very useful to identify fundamental migration properties that cells adopt when confronted with rigid or deformable substrates \cite{angelini2011glass,malinverno2017endocytic}. Nonetheless, the flattened cellular morphology in monolayers gives an unrealistic biological representation that makes impossible to capture the complexity of cell-cell and cell-matrix interactions typical of the three-dimensional (3D) architecture of tumour tissues. Oncological researchers filled this gap by adapting the experimental techniques pioneered by developmental biologists to develop 3D cell aggregates \cite{holtfreter1944study, moscona1952dissociation}. The formation of tumour spheroids  can be achieved by means of different experimental techniques, such as seeding cells on low-adhesion plates, in spinner flasks, or as hanging drops \cite{kunz2004use}.\\
\noindent The increased availability of automated and miniaturized technical assays further pushed the use of 3D cultures for high-throughput screening of the interactions between cancer cells and their micro-environment, highlighting the pivotal importance of chemo-mechanical feedbacks for regulating tumour progression in different experimental conditions \cite{helmlinger1997solid, montel2011stress,puliafito2015three,bubba2018chemotaxis}.
However, how homophilic cell-cell interactions in spheroids of epithelial-derived carcinomas, which represent the vast majority of ‘solid’ tumours, prevail over the cellular interactions with stromal cells, that are in turn essential to promote cancer cell invasion \cite{abercrombie1979contact}, remains by and large unclear. Additionally, spheroids frequently fail in capturing the metastatic potential of malignant cells, that is correlated with their ability to gain a mesenchymal phenotype, to transit from a solid, kinetically arrested to a fluid-like, locomotory state \cite{park2016collective,oswald2017jamming}.
This work builds upon an experimental system that mimics { the spontaneous aggregation of tumoral cells cultered in a monolayer to a clustered morphology, through the formation of flattened thrre-dimensional structures} resembling buds that may emerge during cancer progression. Our aim is to unravel how the chemical and physical cues regulate this topological transition towards an invasive phenotype.

\section*{Experiments and model}
\subsection*{Experiments.} 
\begin{figure*}[ht]
\centering
\includegraphics[width=.85\linewidth]{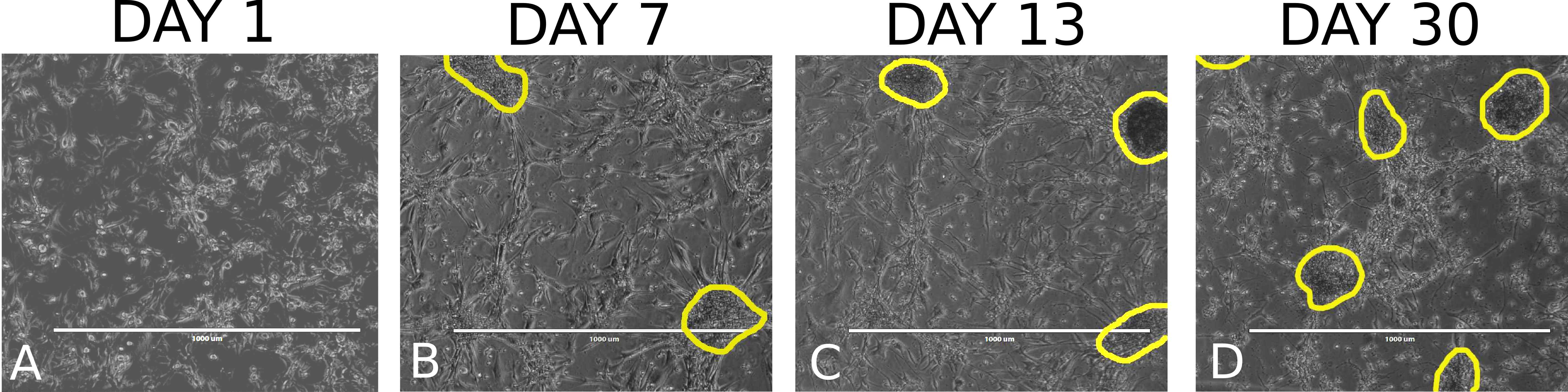}
\caption{ The { phase-contrast} microscope images of the cells in the Petri dish are shown at days 1 (panel A), 7 (B), 13 (C), and 30 (D). The clusters of tumour cells have been identified using an automated procedure of image analysis.  The corresponding contours are traced using a yellow line. The  scalebar is $1$ $mm$. }
\label{fig1}
\end{figure*}

We cultured a glioblastoma cell line (U-87 MG) into commercially available cell culture dishes. { The tumour cells were cultured  under  $37^{o}$C and 5\% ${\rm CO}_2$ in Dulbecco's modified Eagle's medium, with an average initial density $\phi_0$  corresponding to  $43100$ cells/${\rm c}m^2$. Each dish has a diameter of $8.6 \ cm$ and it is filled with a medium layer having a thickness of $0.17 \ cm$. At Day 0 of experiment, cells were seeded and adhered to the plastic bottom, which behaves as an adherent rigid substrate}. At Days 2-3, cells started to re-arrange and to form small adherent clusters. The typical diameter of such aggregates grew over time up to Day 15, when the medium was changed in order to prevent accumulation of waste products.
After an immediate subsequent regression, the {average} clusters domain size continued growing during Days 18-30 with a power-law regime, eventually followed by growth saturation towards a steady state. Cluster formation and growth dynamics were followed using a bright-field microscope for a total duration of 30 days. The observed morphological transition of tumour cells at different stages of their phase ordering dynamics is illustrated in Figure \ref{fig1}. { The three-dimensional structure of the clusters has been investigated using confocal microscopy, showing the onset of tumour aggregates with a flattened morphology, having a maximum  thickness of about  $50$ $\mu m$ (see Supplementary Figure 4). We developed an automated procedure to identify the two-dimensional (2D) area of the tumour clusters from the collected images  using the Fiji software \cite{schindelin2012fiji}. We used library MorphoLibJ \cite{legland2016morpholibj} to quantify the morphological evolution of the clusters over time, computing the area and the  perimeter $p_c$ of each cluster, defining the characteristic length as $l_c =p_c/\pi$.\\ }
From a biological viewpoint, this collective phenomenon of budding closely resembles the self-organization and adaptation mechanisms that are at play in promoting the initial stage of metastasis. Frequently, motile tumour cells migrate by forming adherent clusters that despite a striking degree of phenotypic intercellular heterogeneity maintain a high correlation and coordination in the migration features between neighboring cells  \cite{clark2015modes} \\

\noindent \subsection*{Model.} The clustering dynamics is studied by means of a theoretical model and numerical simulations.
We consider the biological system within the Petri dish {as a 2D incompressible mixture} composed by a cellular phase with volume fraction $\phi_c=\phi_c({\bf x},t)$, such that $0\le\phi_c\le1$,  and a liquid phase with volume fraction $\phi_l= (1-\phi_c)$, so to enforce the overall saturation.  In the following, the subscripts $c,l$ will refer to the cellular and liquid phases, respectively.
{ Motivated by the flattened morphology of the observed clusters, we consider a two dimensional mixture made of two}  phases having about the same density $\rho$ as water, and we impose the following mass balance equations:
 \begin{equation}
 \rho \frac{\partial \phi_i}{\partial t}+\rho  \ \text{div}(\phi_i \mathbf{v}_i)={\Gamma_i} (\phi_i, n), \qquad \quad i=(c,l)
\label{mb}
 \end{equation}
 where  $\mathbf{v}_i$ is the phase velocity, and  ${\Gamma_c}=-{\Gamma_l}\ge0$ is the cellular growth rate, that is assumed to depend on the local concentration of the nutrient field {$n({\bf x},t)$}, i.e. glucose.\\
 \noindent { Beyond  existing sharp-interfase approaches to tumour mechanics \cite{breward2002role,ambrosi2002mechanics,lubkin2002multiphase}}, we consider a diffuse interface model by assuming that the { grand potential or Landau} free energy $E=E(\phi_c,n)$ of the biological system per unit mass is given by:

\begin{equation}
\label{E}
E=\int_{\Omega}\left({\chi}_c\psi(\phi_c)+\frac{\epsilon^2}{2}|\text{grad} (\phi_c)|^2+{\frac{\chi_n}{2} n^2}\right)d\mathbf{x},
\end{equation}

where ${\Omega}$ is the {2D} spatial domain occupied by the dish,  ${\chi}_c \psi(\phi_c)$ is a {  local cell-cell interaction potential of the Lennard-Jones type}, {$\frac{\chi_n}{2} |n|^2$ is the chemical potential} and  $\epsilon^2$ is an energy penalty to prevent the formation of large gradients of tumour concentration. The local interaction described by $\psi(\phi_c)$  {is attractive force for low cellular concentration and  repulsive for $\phi_c>\phi_e$, where $\phi_e$ is a homeostatic concentration such that $\psi^{'}(\phi_e)=0$ \cite{byrne2003modelling}}. Furthermore, we consider a viscous dissipation $W$ between the cells { and the liquid phase}, such that:
\begin{equation}
\label{W}
W=\int_{\Omega}\left(M \ \phi_c \ (\mathbf{v}_l-\mathbf{v}_c)^2\right)d\mathbf{x},
\end{equation}
 where $M$ is a friction parameter, { that is  determined by the cell-substrate interaction}.

{ The Onsager  variational principle \cite{doi2011onsager}} is used to derive the evolution equation for the biological system as the minimizing solutions of the  Rayleighian functional $(W+dE/dt)$ with respect to the phase velocities \cite{chatelain2011emergence}. { In contrast to previous diffuse inteface approaches to tumour growth modeling \cite{lowengrub,wise}, we remark that we use here a physically motivated single-well potential $\psi$
with  a logarithmic singularity as $\phi_c\rightarrow 1$,  and a degenerate motility in $\phi_c=0$.}

 { Assuming a  a viscous--dominated  growth regime}, we obtain the following  fourth--order evolutionary partial differential equation:
 \begin{equation}
\label{CH}
\frac{\partial \phi_c}{\partial t}- \text{div}\biggl(D(\phi_c) \text{grad} (\mu)\biggr)=\Gamma_c (\phi_c, n)/\rho,
\end{equation}
where $D(\phi_c)= {\phi_c(1-\phi_c)^2}/{M}$, $\mu=\delta E/\delta \phi_c= ({\chi}_c \psi'(\phi_c)-\epsilon^2 \nabla^2 \phi_c)$ is the mechanical potential, where $\nabla^2$ denotes the Laplacian operator. The nutrient dynamics is finally given by the following reaction-diffusion equation:
  \begin{equation}
\label{n}
\frac{\partial n}{\partial t}=D_n \nabla^2 n + \Gamma_n (\phi_c, n),
\end{equation}
where $D_n$ is the diffusion coefficient of the chemical concentration and  $\Gamma_n$ is the nutrient growth rate.  Assuming linear growth laws both for the tumour cells and the nutrient,  we consider $\Gamma_c= \rho \gamma_c \phi_c \biggl({n}/{n_s}-\delta_c\biggr)$, where $\gamma_c$ is the proliferation rate of the cell line, $\delta_c$ is its apoptotic rate, so that $\delta_c \ n_s$ is the threshold glucose concentration for triggering cell death by starvation. Finally, $\Gamma_n= S_n(n_s-n)-\delta_n \phi_c n$, { where $S_n$ is the tumour growth rate}, and $\delta_n$ is the uptake rate by the tumour cells. { Since the thickness of the medium layer inside the dish is much bigger than the thickness of the cell layer, we remark that $S_n$ represents the  averaged transverse out-of-plane influx  that tend to restore the initial value $n_s$, that is obtained as the lubrication limit of the three-dimensional model\cite{ba13}}. { The physiological range of values of the model parameters is given in Supplementary Table 1}. The partial differential system given by  Eqs.(\ref{CH}, \ref{n}) is rewritten in dimensionless form and numerically solved inside a circular domain representing the Petri dish,  enforcing that $\mu$, $\phi_c$ and $n$ { satisfy no-flux boundary conditions} at the boundary. In particular we choose the characteristic time- and length- scales as $t_c=\gamma_c^{-1}$ and $l_n= \sqrt{D_n/\delta_n}$, respectively, {where $l_n$ is the nutrient penetration depth.
}

\section*{Results}
Numerical simulations have been performed by implementing a finite element code in FreeFem++ \cite{hecht2012new}, using linear elements in space and a backward Euler scheme in time,  implementing the algorithm described in the Methods section and resumed in Supplementary Note 1. The cells are seeded homogeneously at an average volume fraction $\phi_0<\phi_e$ such that $\psi^{''}(\phi_0)<0$,{ so that the mixture is initially in a metastable state}.  Accordingly, the biological system immediately undergoes a spinodal decomposition, with the tendency to create spatial domains that are pure in each phase. The typical width of the diffuse interface separating  such domains is expected of the order of the single cell size, i.e.  $d=\epsilon/\sqrt{\chi_c}\simeq 10 \mu m$. Since the mechanical potential $\mu$ only contains a short-range interaction potential between cells, small variations of the initial conditions on $\phi_c$ are also irrelevant to the long-time phase-ordering dynamics. {Thus, we set:
\begin{equation}
\langle \phi_c({\bf x},0) \phi_c({\bf x}^{'},0)\rangle = 0.002 \ \delta({\bf x}-{\bf x}^{'}),
\label{incond}
\end{equation}
where the notation $\langle \cdot \rangle$ indicates the average over an ensemble of initial conditions},
 in order to evaluate the late-time evolution of the deterministic Eqs.(\ref{CH},\ref{n}) subject to white-noise fluctuations of the initial condition around $\phi_0$. { In particular, the numerical simulations in the following have been perfomed using an ensemble of six different initial conditions following \eqref{incond}.}
 {Here $\delta(\cdot)$ is the Dirac distribution function, and the pre-factor $0.002$ represents the size of the fluctuations}. {In order to describe the change of medium, we impose $n\equiv 1$  at day $15$ in our simulations.}
 { Using a sensitivity analysis  on the model parameters in the physiological range  (see Supplementary Note 2),   we found that the spontaneous aggregation of tumour cells is mostly controlled} by the dimensionless parameter $\Pi_d= (\chi_c \ t_c)/(M l_n^2)\simeq {24.2}$, highlighting that { the cell-cell adhesion forces must be of the same order as the viscous  drag forces  exerted as the growing cellular phase expands over the substrate}. The results of the numerical simulation are displayed in Figure 2 in qualitative comparison with the images collected during the in-vitro experiments. A full simulation is also available as Supplementary Movie 1.

   \begin{figure*}[th!]
\centering
\includegraphics[width=.95\linewidth]{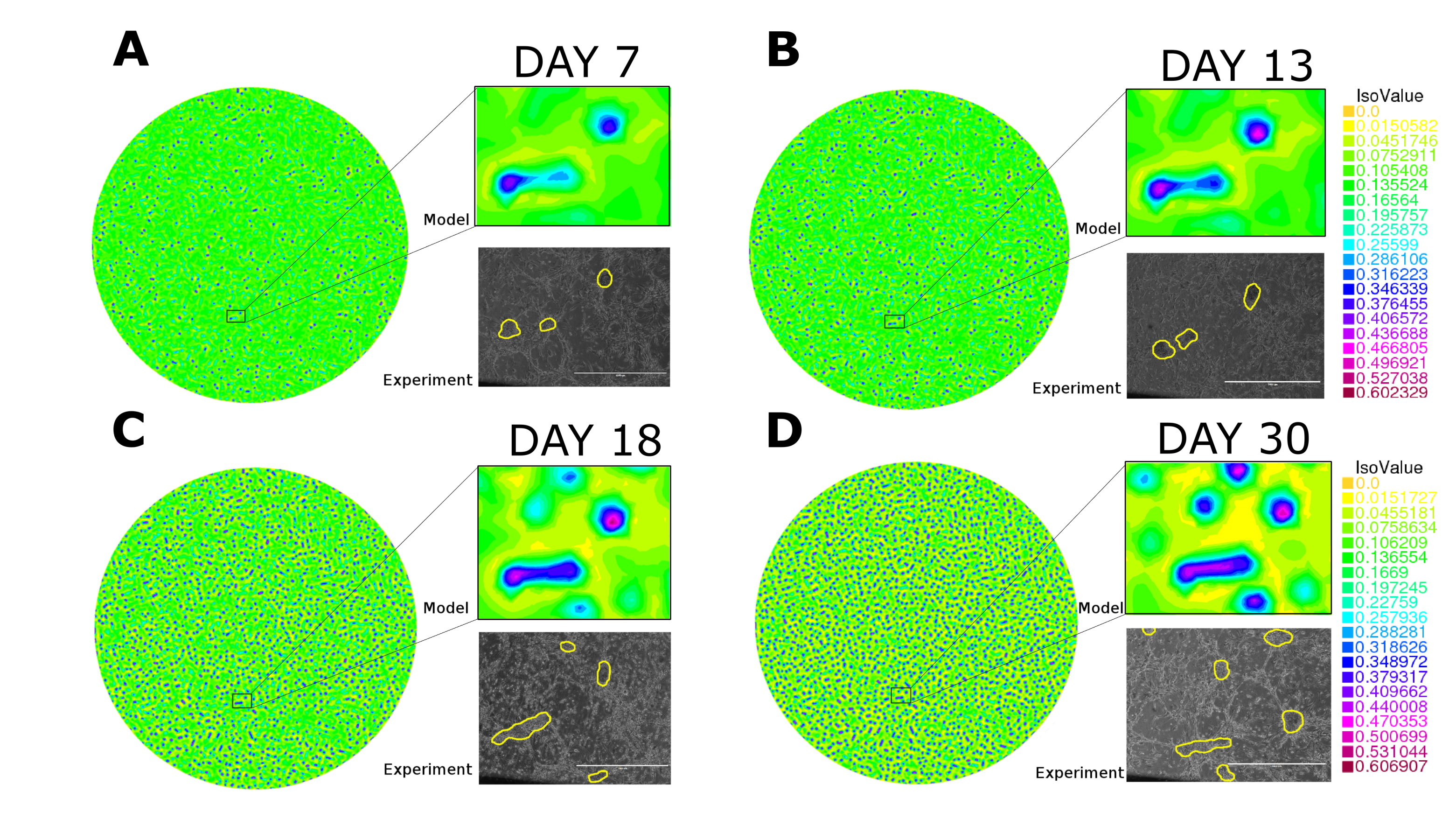}
\caption{Spatial distribution of the tumour volume fraction $\phi_c$ is depicted at day 7 (panel A), day 13 (B), day 18 (C) and day 30 (D) using a color scalebar. The circular domain of the numerical simulation represents the Petri dish, having a diameter of $8.6$ $cm$.  { The insets show a qualitative comparison between the model results (top) and the experiments (bottom).  The dimensionless parameters are indicated in Table 1. The  scalebar in the experimental images is $1$ $mm$, the color scale refers to the tumour volume fraction.}}
\label{numsol}
\end{figure*}

The far-from-equilibrium kinetics of phase ordering of tumour cells via the spontaneous budding is studied using  statistical mechanics tools to highlight its universal features.
{ Following \cite{bray2002theory}  we make the fundamental scaling hypothesis that the tumour clusters  follow a time-independent growth law that depends on} a single characteristic length $L(t)$. Accordingly,  the { two--point} equal time correlation function {$C({\bf r},\phi_c({\bf x},t))$} is invariant after re-scaling lengths by $L(t)$ \cite{bray2002theory}, i.e:
{
\begin{equation}
C({\bf r},\phi_c({\bf x},t))=\langle \phi_c({\bf x}+{\bf r},t) \phi_c({\bf x},t)\rangle= f\left(\frac{|{\bf r}|}{L(t)}\right).
\label{C}
\end{equation}
}
{ This equal-time correlation function is a nonequilibrium quantity as domain growth is a nonequilibrium process, with short-distance singularities representing sharp interfaces (i.e. defects) in the phase-ordering system.} In particular, the scalar function $f\left({|{\bf r}|}/{L(t)}\right)\sim (-\left({|{\bf r}|})/{L(t)}\right)$  for $d\ll|{\bf r}| \ll L(t)$, since it corresponds to the probability to draw randomly a segment of length $|{\bf r}|$ that intersects an interface in a typical domain of size $L(t)$. Eq.\eqref{C} also implies that the domain size distribution {$\hat{n}(l_c,t)$} of  clusters having typical size $l_c$ {at time $t$ follows a  time-independent master curve} . Thus, the total number $N(\ell,t)$ of clusters with size {less than or equal to} $\ell$ scales as {$N(\ell,t)=\int_0^{\ell} ds \ \hat{n}(s,t) \sim L(t)^{-2}$}. The experimental data highlights that the transition to  self-organization during the  tumour cell evolution occurs at day 11 of in-vitro experiments. The onset of { time-independence} in the curves $N(\ell, t) \ L(t)^{-2}$ versus the scaled length $\ell/L(t)$ is  depicted in Figure 3 (A), where $L(t)$ is calculated as the { mean value of the clusters size $l_c$ measured in experiments  at time $t$}.\\

 \begin{figure}[th!]
\centering
\includegraphics[width=.45\linewidth]{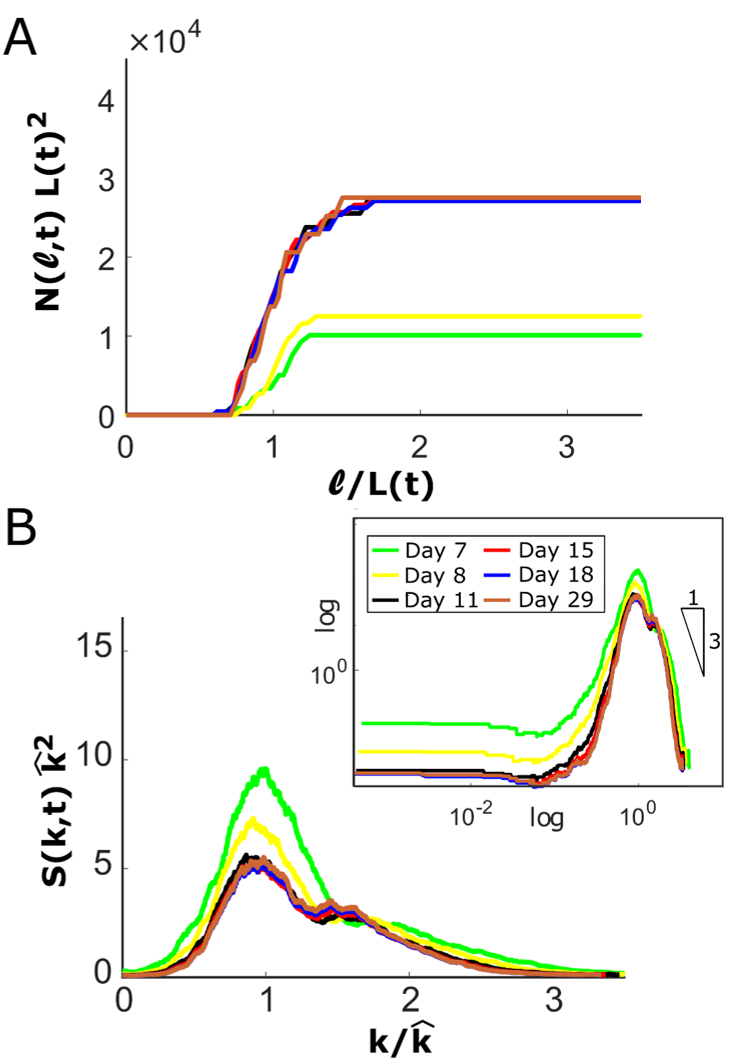}
\caption{ (A) The scaled cluster distribution function $N(\ell,t)L(t)^2$ is calculated from experimental images and depicted over the scaled length $\ell/L(t)$. (B) The scaled structure function $S({\bf k},t)\hat{k}^2$  is calculated from  numerical simulation and depicted over the scaled length $k/\hat{k}$. The inset shown a log-log plot highlighting a characteristic regime given by the Porod law. A statistical correlation of the tumour clustering is shown by the collapse on master curves starting since day $11$.}
\label{fig3}
\end{figure}

An identical transition to a  { universal } clustering regime is  observed in numerical simulations by calculating the structure function $S({ k},t)$, which is the spherically averaged Fourier transform of $C({\bf r},\phi_c({\bf x},t))$. Since the characteristic length is given by $L(t)=\hat{k} ^{-1}$, {where $\hat{k}$ is the first moment of $k$ with respect to the spherically averaged structure function (see the Methods section for details)}, we found that  $S({\bf k},t) \hat{k}^{2}= g(\hat{k}/k)$ after day $11$, as depicted in Figure 3 (B). Recalling that  $f\left({|{\bf r}|}/{L(t)}\right)\sim (-\left({|{\bf r}|})/{L(t)}\right)$ for $L^{-1}<<k<<d^{-1}$, we recover the universal tail of the master curve $g(\hat{k}/k)\sim (\hat{k}/k)^{-3}$, known as the Porod law \cite{porod1951}.\\
{We note that  the no-flux  boundary conditions make the homogeneous scaling hypothesis \eqref{C} invalid in proximity to the dish boundary. Thus   the scaling laws of the correlation function and of the structure function do not hold over a distance smaller than the typical length scale of the clusters from the boundary. Since in our simulations we are resolving  up to the average length--scale of the clusters, taking a mesh size in the range $0.44 - 1.17 \ l_n$ (see the Methods Section for details), we  resolve the kinetics only up to the beginning of the Porod law regime.}
\noindent {The evolution over time of the average size of the tumour aggregates}  is illustrated in Figure $4$, displaying the quantitative agreement between the characteristic cluster size $L(t)$ during in--vitro experiments and in the numerical simulations.  The tumour clusters enter into a coarsening phase at day $11$, that is followed by a metastable damped oscillatory regime driven by the kinetic coupling of the glucose concentration. { After the medium change at day $15$}, the phase ordering dynamics initially displays a universal growth law of the cluster size followed by a saturation regime. We also find a good quantitative agreement between the overall number of clusters appearing over time  in numerical simulations versus the experimental measures {(see Supplementary Figure 1)}.
  \begin{figure}[ht!]
\centering
\includegraphics[width=.45\linewidth]{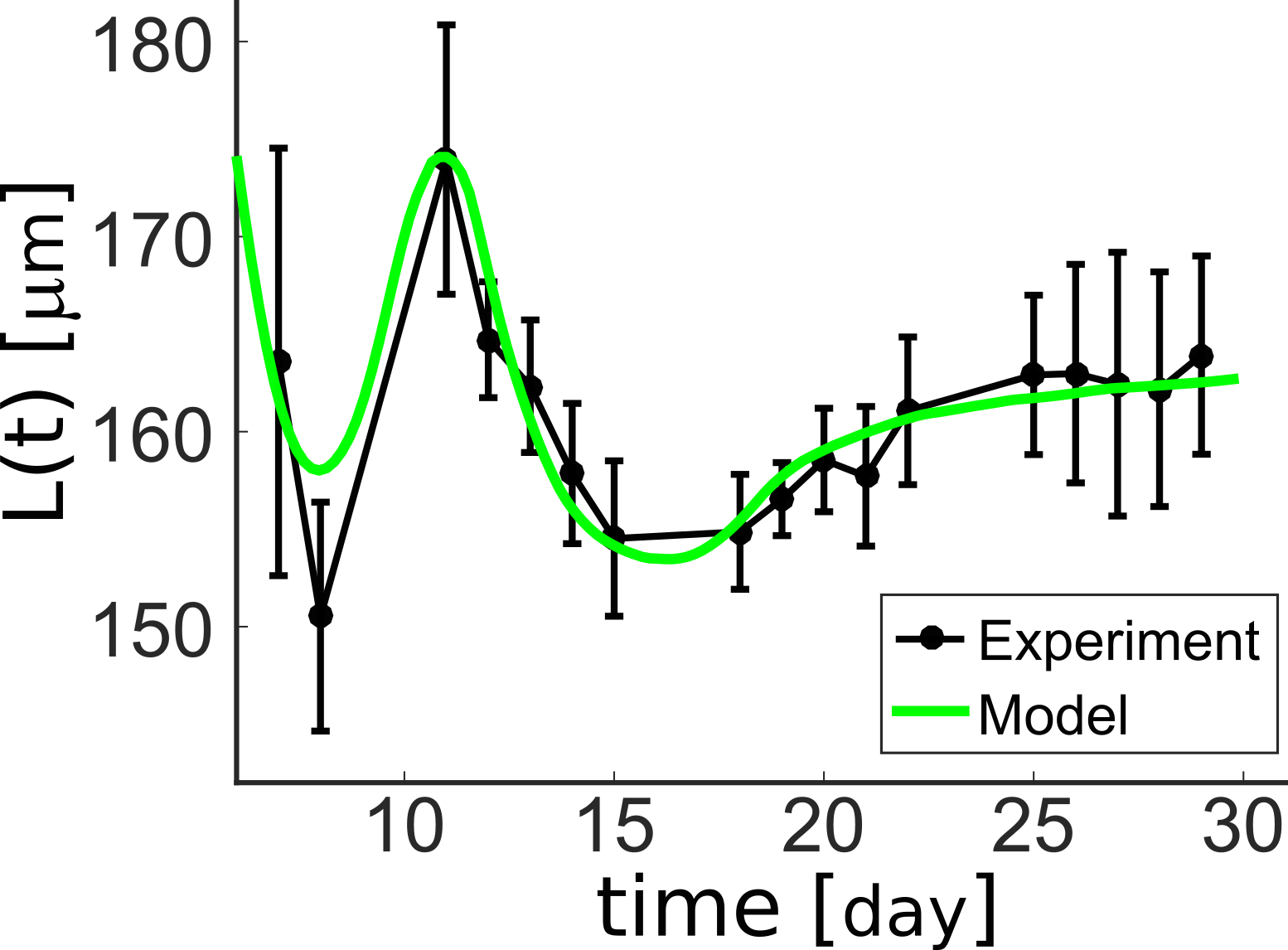}
\caption{Characteristic length $L(t)$ of the tumour clusters over time. The green solid line refers to the numerical simulation, whilst the black markers indicate the experimental measures. The bars denote the error.}
\label{fig4}
\end{figure}
The characteristic increase of the average size of the tumour aggregates over time is ruled  by a power law $L(t)\sim t^\alpha$, with $\alpha \simeq 0.22$ in the coarsening phase (see Supplementary Figure 2). The  exponent $\alpha$ is found to be bounded by two characteristic regimes. { The upper limit $\alpha=1/4$ corresponds to the universal dynamics imposed by the phase-dependent mobility\cite{mazenko1990theory} , since the surface diffusion is enhanced in the vicinity of the interfaces, where the mobility is suppresed  \cite{ohta1988ordering}}.  The lower limit $\alpha=1/5$  characterizes the kinetic transitions dominated by a cluster reaction and diffusion process \cite{puri1997phase}.  { The  stochastic transport of tumour cells determines a  diffusivity $D_s$ that is inversely proportional to the cluster surface (order $L^2$) multiplied by the displacement of the cluster center of mass (order L), so that $D_s \sim L^{-3}$}, and $t \sim L^{2}/D_s \sim L^{5}$ \cite{binder1974theory}. During coarsening, the numerical simulations also reproduce the experimental  merging dynamics between two small clusters, as shown in the insets of Figure 2. In physical terms, { their short--range interaction potential is domitated by the exponential tail of the domain walls, being  of the order $\sim exp(-|{\bf r}|/d)$. This effect  creates} an attractive force dipole that eventually drives the merging of the two interfaces, that is favoured by the low energy cost of surface diffusion \cite{rutenberg1994phase}.
Finally, the later saturation growth regime is dictated by the chemical freezing of coarsening when the cluster size $L(t)$ reaches a saturation length which is of the order of half the glucose penetration length $l_n/2\simeq 170\  \mu m$. This can be understood by considering that when both typical lengths are of the same order, the steady state solution of Eqs.(\ref{CH}, \ref{n}) can be stabilized { by a Turing--like mechanism\cite{carati1997chemical} }. Thus, the reaction-diffusion dynamics eventually selects a steady pattern in phase-ordering tumour clusters, and the transition time from coarsening to saturation is of the order $t \sim l_n^{1/\alpha}$.

\section*{Discussion}

This work demonstrates that the spontaneous \textit{in vitro} aggregation of glioblastoma cells from a monolayer to ordered clusters is a self-organised, non-equilibrium phenomenon, occurring when the cell-to-cell adhesive interaction is of the same order as the  { drag forces exchanged whilst cells  expand over the rigid substrate. The results of our continous multiphase model are consistent with those obtained using agent-based approaches \cite{agnew2014distinguishing,cui2016mechanistic}, highligthing that the proliferation rate of the cells and its mechanical interaction with the surrounding environment  are  dominant mechanisms in triggering cell aggregation.  Notably, this peculiar process was earlier reported in U-87 MG cells\cite{yu,ledur}, but never characterized in depth}.

 {Our results confirm that tumour aggregates could undergo drastic rearrangements in their mechanical properties, by tuning cell-cell and cell-substrate adhesive properties. This capacity constitutes a pre-requisite for the alteration of epithelial polarity, that in turn is necessary to promote invasion to surrounding tissues and subsequent dissemination to distant sites \cite{pease2012spontaneous}. Spontaneous clustering might be a more general property of cancer cells, paving the way for the successive acquisition of epithelial-mesenchymal transition (EMT)-like features, independently of tumour genetic and molecular heterogeneity. Indeed, this topological switch is found, here, to be controlled by mechanical and chemical cues without a full and overt change in the cell identity. }

    In conclusion, this works unravels a new universality class governing the spontaneous aggregation of tumour cell populations. Small random perturbations driven by genetic instability and phenotypic heterogeneity may drive configurational changes to unstable disorder states that macroscopically emerge as complex spatiotemporal patterns that are statistically correlated. These new insights on the { universality} in the kinetic evolution of spontaneous tumour clustering represents an important step for improving our ability to develop predictive models of pattern formation in cancer, that will be a key enabling technology for developing new strategies to control the complexity of cancer invasion.

\section*{Methods}
\footnotesize
\subsection*{Cell cultures.}
  {The experiments were performed at IFOM laboratory. A glioblastoma cell line, U-87 MG, was obtained by ATCC (American Type Culture Collection)
 consortium. Cells were cultured at room temperature of $37^{o}C$ and 5\% $CO_2$ in Dulbecco’s modified Eagle’s medium { plus Fetal Bovine Serum (10\% in volume), L-Glutamine (2 mM), Sodium Pyruvate (1 mM), Non-Essential Amino-Acids (1\% in volume) and Penicillin/Streptomycin (100$ \mu$g/ml). At Day 0, about 2.5 millions of U-87 MG cells were seeded into two dishes (considered as technical replicas) with a total medium volume of 10 ml/each. Medium was changed at Day 15, to prevent accumulation of waste products}.}\\

\subsection*{Image and data analysis.} {The formation and growth of clusters of U-87 MG cells were followed by {phase--contrast} microscope. Images were acquired at $4X$ magnification using an EVOS-FL Imaging System (ThermoFisher Scientific), equipped with a high-sensitivity monochrome camera (1360x1024 pixels, 6.45$\mu m$/pixel). A reference grid was drawn on the bottom surface of each  plastic dish in order to follow the spatial distribution of clustering over time. Twelve different fields for each dish were randomly-chosen within the grid and imaged five days a week for $30$ days, starting from cell seeding. The dimension of each field was $2.2\times 1.6$ $mm^2$.\\
\noindent The collected images were analyzed with Fiji software\cite{schindelin2012fiji}  and its library MorphoLibJ\cite{legland2016morpholibj} to quantify the  clusters morphology and its growth over time.\\
\noindent  The automated procedure consists in the following steps:
\begin{itemize}
\item \textit{De-noising} of each  8-bit gray-scale image using  a (local) \textit{minimum filter}  followed by an histogram stretching operation by pixels saturation in order to enhance contrast;
\item \textit{Morphological reconstruction} applying a (non-local) \textit{extended  minima filter} first to the de-noised image and later to the reconstructed image. The tolerance of the  filters is automatically set as half of the width of the pixel histogram;
\item \textit{Filtering} using \textit{gray--scale attribute}  to eliminate clusters smaller than 10 cells,  and \textit{opening operation} to smooth  the cluster boundary and to remove boundary oscillations of the same order as the cell diameter;
\item  \textit{Morphological segmentation} using a watershed algorithm to detect the  cluster domains, and to compute  the area $A_c$, the perimeter $p_c$, and the geodetic diameter of each cluster. We also compute the circularity index of each cluster, defined as $4\pi A_c/p_c^2$.
\end{itemize}
An illustrative  sketch of the steps of this automated procedure is given in Supplementary Figure 3.\\
{ For confocal analysis of clusters, { that is depicted in Supplementary Figure 4], U87-MG cells were seeded on a $\mu$-Slide 8 Well (Ibidi, code 80821) at the same density of previous experiment. At Day 7 from cell seeding, cell nuclei were stained with NucBlue Live ReadyProbes reagent (ThermoFisher, code R37605) and observed with a Leica SP8 confocal microscope. Images were acquired with a 20X objective N.A. 0.75, with an additional 1.9 zoom factor. Vertical sections of clusters were obtained by performing an xzy acquisition. Nuclear staining was imaged after excitation with 405 laser line, whereas coverslip surface was imaged in reflection mode with 488 laser line.}}

\vspace{1cm}

\subsection*{Dimensionless analysis.} {For the ease of notation, in this Section we indicate the cellular volume fraction through the variable $\phi$, dropping the subscript $c$}. We introduced the following dimensionless variables:
{\begin{equation}
\label{eqn:16}
\hat{x}=\frac{x}{l_n}, \quad \hat{t}=t\gamma_c, \quad \hat{n}=\frac{n}{n_s}.
\end{equation}}
Accordingly, the governing equations in Eqs.(4, 5) of the article can be rewritten in dimensionless form as:
{ \begin{equation}
\label{eqn:17}
\begin{cases}
\frac{\partial \phi}{\partial \hat{t}}- \Pi_d {\hat{\text{d{i}v}}}(\phi(1-\phi)^2\hat{\text grad} \hat{\mu})=\phi (\hat{n}-\delta_c),\\
\hat{\mu}=\psi'(\phi)-\gamma^2\Delta \phi,\\
\frac{1}{v}\frac{\partial n}{\partial \hat{t}}=\hat{\Delta} \hat{n}-\phi \hat{n} + \beta(1-n),
\end{cases}
\end{equation}}
where the hatted operators are also dimensionless. { The cell-cell interaction potential is given by :\\

$$ \psi(\phi)= -(1-\phi_e)\left(\log(1-\phi)+\phi+\frac{\phi^2}{2}\right) -\frac{\phi^3}{3} $$}\\

Thus, the evolution of the biological system is governed by four dimensionless parameters:
{ \begin{equation}
\label{eqn:18}
\Pi_d=\frac{\chi_c t_c}{M l_n^2}, \quad \gamma^2=\frac{\epsilon^2}{\chi_c l_n^2}, \quad {v}=\frac{\delta_n}{\gamma_c}, \quad \beta=\frac{S_n}{\delta_n}.
\end{equation}}
whose values  are collected in  Table 1.

\begin{table}[h]
\caption{Values of the dimensionless parameters of the model}
\label{tab:1}    
\begin{center}
\begin{tabular}{|llll|}
\hline
Parameters &  Formula or Ref. & Physical range & Selected value \\
\hline
{$\phi_e$} & \cite{jain} & [0.6 - 0.87] & 0.6   \\
{$\phi_0$} & $\pi (d/2)^243100\phi_e$ & [0.18 - 0.1853] & 0.18   \\
{$\delta_c$} & \cite{bedogni} & [0.1 - 0.33] & 0.3   \\
$\Pi_D$ & $\frac{\chi_c}{M\gamma_c l_n^2}$ & [0.3 - 902] & 24.2   \\
 $\beta$ & $S_n/\delta_n$& $[2.52 \, 10^{-4}, 0.0639]$  & 0.045 \\
$\gamma$ & $\epsilon/(\sqrt{\chi_c}l_n)$ & [0.0116 - 0.5556] & 0.027   \\
 $v$ & $\delta_n/\gamma_c$ & [28.0183 - 7056.4114]& 40.19 \\
 \hline
\end{tabular}
\end{center}
\end{table}

\subsection*{Numerical method.}
We build an unstructured  mesh of 345.653 elements with a mesh size in the range $0.44 - 1.17 \ l_n$.  Given a partition of the time interval $[0,T]$ into {$M$} discrete intervals with endpoints {$t_m=m\Delta t$, $m=1,\dots,M$, $\Delta t=T/M$}, the time derivative of the field $\phi(t)$ at time $t_{m}$ is approximated as the finite difference ratio $(\phi(t_m)-\phi(t_{m-1}))/\Delta t$. We set $\Delta t=0.5 (d/l_c)^2$ as a suitable time scale for the phase ordering dynamics.\\
Let $J$ be the set of nodes of $\mathcal{T}_{h}$ and $\{x_j\}_{j\in J}$ be the set of their coordinates. Moreover, let $\{\chi_j\}_{j\in J}$ be the Lagrangian basis functions associated with each node $j\in J$.
Denoting by $\pi^h:C(\bar{\Omega})\rightarrow S^h$ the standard Lagrangian interpolation operator we define the lumped scalar product as
{ \begin{equation}
\label{eqn:lump}
(\eta_1,\eta_2)^h=\int_{\Omega}\pi^h(\eta_1(x)\eta_2(x))dx\equiv \sum_{j\in J}(1,\chi_j)\eta_1(x_j)\eta_2(x_j),
\end{equation}}
for all $\eta_1,\eta_2\in C(\bar{\Omega})$.
The $\mathcal{L}^2$ lumped projection operator $\hat{P}^h:\mathcal{L}^2(\Omega)\rightarrow S^h$ is defined by
{ \begin{equation}
\label{eqn:l2proj}
 (\hat{P}^h\eta,\chi)^h=(\eta,\chi) \quad \forall \chi \in S^h.
\end{equation}}
Starting from a datum $\phi_{0},n_{0}\in H^{1}(\Omega)$ and $\phi_{h}^{0}=\hat{P}^{h}\phi_{0}$, $n_{h}^{0}=\hat{P}^{h}n_{0}$, with $0\leq \phi_{h}^{0}<1$ and $0< n_{h}^{0}<1$, we consider the following fully discretized problem:\\

\noindent \textit{Problem $\mathbf{P}^{h}$.}
\label{pbm:ph0}
For $m=1,\dots,M$, given $\phi_{h}^{m-1}\in K^{h}$, find $(\phi_{h}^{m},\mu_{h}^{m},n_h^m)\in K^{h}\times S^{h}\times K^{h}$ such that for all $(v_h,w_h,q_h)\in S^h\times K^h\times S^{h}$,
{
\begin{equation}
\label{pbm:ph}
\begin{cases}
\displaystyle \biggl(\frac{\phi_{h}^m-\phi_{h}^{m-1}}{\Delta t},v_h \biggr)^h+\Pi_d(b(\phi_{h}^{m-1})\nabla \mu_{h}^m,\nabla v_h)=(\phi_{h}^{m-1}(n_h^m-\delta_c),v_h)^h,
\\
\displaystyle \gamma^2(\nabla \phi_{h}^m,\nabla(w_h -\phi_{h}^m))+(\psi'_{1}(\phi_{h}^m),w_h -\phi_{h}^m)^h\geq (\mu_{h}^m-\psi'_{2}(\phi_{h}^{m-1}),w_h -\phi_{h}^m)^h,\\
\displaystyle \biggl(\frac{n_{h}^m-n_{h}^{m-1}}{\Delta t},q_h \biggr)^h+v(\nabla n_{h}^m,\nabla q_h)=v\beta (1-n_h^m,q_h)^h-v(\phi_{h}^{m-1}n_h^m,q_h)^h,
\end{cases}
\end{equation}
}
where the following convex splitting of the potential:
{
\begin{equation}
\label{eqn:convsplit}
\psi_1(\phi)=-(1-\phi_e)\ln(1-\phi), \quad \psi_2(\phi)=-\biggl[\frac{1}{3}\phi^3+\frac{1-\phi_e}{2}\phi^2+(1-\phi_e)\phi \biggr]
\end{equation}
}
is considered to ensure the gradient stability of the numerical approximation.
The variational inequality in \eqref{pbm:ph} enforces the positivity of the discrete solution since it implies that $\phi_h^m$ is projected onto the space with positive values $K^h$. The lumping approximation of the $\mathcal{L}^2$ scalar product introduced at the left hand side of \eqref{pbm:ph} allows the discrete solution to select the physical solutions with compact support and moving boundary from the ones with fixed support, as described in \cite{elliott1996cahn}.
Moreover, we introduce a proper subdivision of $J$ into a set of \textit{passive nodes} $J_0(\phi_h^{m-1})\subset J$ such that $ (\phi_h^{m-1},\varphi_j)=0$ for $j\in J_0(\phi_h^{m-1})$
 and a set of \textit{active nodes} $J_+(\phi_h^{m-1})=J \setminus J_0(\phi_h^{m-1})$. A passive node is thus characterized by the fact that $\phi_h^{m-1}\equiv 0$ on the support of the basis function associated to it.\\

Finally, we solve the system \eqref{pbm:ph} for every time step by  generalising the iterative procedure introduced in  \cite{Barrett1999,agosti2017cahn} to formulate an algorithm, in which the variational inequality  is solved by a null space method on the active nodes, whereas we impose the constraint $\phi_h^{m}\equiv 0$ on the passive nodes. The algorithm is described in Supplementary Note 1.

{At day $15$ we impose $n_h^m\equiv 1$ in our simulations in order to
describe the change of medium. In order to tune the parameters of the model against the experimental results, we performed different simulations starting from the values $\phi_e = 0.6$, $\phi_0 = 0.18$, $\delta_c=0.3$, and varying the parameters $\Pi_d$ and $\beta$ on a grid of discrete values in order
to minimize the $\mathcal{L}^2$ norm of the difference
between the simulation and the experimental profiles of $L(t)$. The set of parameters which gave the least distance is reported in Table \ref{tab:1}.}

\subsection*{Structure factors in simulations.} {Let {$N_v=L_{\Omega}^2$ be the total number of nodes in the mesh, $L_{\Omega}$ is the linear size of the mesh lattice, and $<\phi_h^m>_{\bf x}$ is the spatial average of $\phi_h^m$}. The spherically averaged structure factor $s(k,t)$ is calculated as the average of the structure factor $S(\mathbf{k},t)$ over all wave--vectors of magnitude $(k-\Delta k)$ and $(k+\Delta k)$, being:
{
{\begin{equation}
\label{eqn:sktscalingstructure}
S(\mathbf{k},m\Delta t)=
\biggl<\frac{1}{N_v}\biggl|\sum_{\mathbf{r}}e^{-i\mathbf{k}\cdot \mathbf{r}}[\phi_h^m(\mathbf{r},t)-<\phi_h^m>_{\bf x}]\biggr|^2\biggr>,
\end{equation}}
}
where  the summation  runs over the lattice points.  The summation over $\mathbf{r}$ in Equation \eqref{eqn:sktscalingstructure} is calculated as a Fourier discrete transform, with {$\mathbf{k}=2\pi \mathbf{n}_v/L_{\Omega}$}, where the vector {$\mathbf{n}_v=[n_{v,1},n_{v,2}]$, with $n_{v,1},n_{v,2}$ spanning from 0 to $\sqrt{N_v}-1$}, indicates the position in the dual lattice. Since we are starting with a field $\phi_h^m$ over a circular domain  of the Petri dish, we use a conformal mapping from the circular to a rectangular domain by the coordinate transformation $x=r \cos \theta, y=r \sin \theta$, and then we calculate the Fourier discrete transform of $\phi_h^m(r,\theta)$ over the Cartesian domain $[0,R]\times [0,2\pi]$. The normalized spherically averaged time dependent structure factor $S(k,t)$  reads
{
$S(k,t=m\Delta t)={s(k,t)}/{(<(\phi_h^m)^2>_{\bf x}-<\phi_h^m>_{\bf x}^2)}$, so that $L(t)=\hat{k}^{-1}:=\bigl({\sum_{k}k \ s(k,t)}/{\sum_{k}s(k,t)}\bigr)^{-1}$.}}\\


\section*{Acknowledgements}
 This work was supported by Associazione Italiana per la Ricerca sul Cancro (AIRC) through the MFAG grant 17412. The authors are indebted to L. Casarico for the kind assistance in  experiments. P.C. acknowledges the support of Les Treilles Foundation during the residential study "Mathematical modeling of growth and tissue repair".

\section*{Author contributions statement}

P.C. and G.S. conceived the study, P.C and A.A. performed the theoretical and the numerical analysis, S.M. and G.S. conceived and performed the experiments, all authors contributed in writing and reviewing the manuscript.

\subsection*{Data Availability}
The authors declare that all data supporting the findings of this study are available within the article and its Supplementary Information, or are available from the authors upon request.\\

\subsection*{Competing interests}
 The authors declare no competing  interests.\\

\bibliography{prlbib}

\end{document}